\documentclass[twocolumn,preprintnumbers,amsmath,amssymb]{revtex4}

\usepackage{graphicx}
\usepackage{dcolumn}
\usepackage{bm}

\begin{document}

\title{\bf DFT-GGA errors in NO chemisorption energies on (111) transition metal surfaces:  Possible origins and correction schemes}

\author{Xu Huang and Sara E. Mason}
\affiliation{ Department of Chemistry\\ University of Iowa, Iowa City, Iowa 52242}%

\date{\today}

\begin{abstract}
Here we investigate whether well-known DFT-GGA errors in predicting
the chemisorption energy ($E_{\rm chem}$) of CO on transition metal
surfaces manifest in analogous NO chemisorption systems.  To verify
the occurrence of DFT-GGA overestimation of the back-donation
mechanism in NO chemisorption, we use electronic structure analysis to
show that the partially filled molecular NO 2$\pi^{*}$ orbital
rehybridizes with the transition metal $d$-band to form new bonding
and anti-bonding states. We relate the back-donation charge transfer
associated with chemisorption to the promotion of an electron from the
5$\sigma$ orbital to the 2$\pi^{*}$ orbital in the gas-phase NO
G$^{2}\Sigma^{-}\leftarrow \rm{X}^{2}\Pi$ excitation.
We establish linear relationships between $E_{\rm chem}$ and $\Delta
E_{\rm G\leftarrow X}$ and go on to formulate an $E_{\rm chem}$
correction scheme in the style of Mason {\it et al.}, {[Physical
    Review B {\bf 69}, 161401(R)]}.  We apply the NO $E_{\rm chem}$
correction method to the (111) surfaces of Pt, Pd, Rh, and Ir, with NO
chemisorption modeled at a coverage of 0.25 ML.  We note that both the
slope of $E_{\rm chem}$ {\it vs.}  $\Delta E_{\rm G\leftarrow X}$ and
the dipole moment depend strongly on adsorption site for each metal,
and we use this fact to construct an approximate correction scheme which
we go on to test using NO/Pt(100) chemisorption.
\end{abstract}


\maketitle
\section{\label{sec:level1}Introduction }

The chemisorption of nitric oxide (NO) and carbon monoxide (CO) on
transition metal surfaces has drawn much research attention due to the
fundamental importance of these processes to applications in
catalysis.  In this study, we focus on the issue of density functional
theory (DFT) predictions of the chemisorption energy, $E_{\rm chem}$,
of NO on transition metal surfaces.  In particular, we investigate
whether well-known shortcomings in DFT predictions for $E_{\rm chem}$
of CO manifest in NO/metal systems.  This so-called ``CO/metal
puzzle''~\cite{Feibelman01p4018} can be summarized as the tendency for
generalized gradient approximation (GGA) functionals (such as
PW91~\cite{Perdew92p6671}, PBE~\cite{Perdew96p3865} and
RPBE~\cite{Hammer99p7413}) to place the empty CO 2$\pi$* orbital
too low in energy relative to the metal $d$-band.  This results in an
unrealistic strengthening of the 2$\pi$*- $d$-band interaction, which
is reflected in an overestimation of the chemisorption bond strength.

Further insight into the CO/metal puzzle can be obtained by
considering the Hammer-Morikawa-Norskov (HMN) model for the $d$-band
contribution to CO $E_{\rm chem}$~\cite{Hammer96p2141}.  The back-donation term, $E_{\rm chem}^{d \rightarrow 2\pi^*}$, thought
to dominate chemisorption on late transition metal
surfaces, is given by:

\begin{eqnarray}
 E_{\rm chem}^{d \rightarrow 2\pi^*} = \frac {-4 f V_{\pi}^2}
 {(\epsilon_{2\pi^*}-\epsilon_d)}
\end{eqnarray}

\noindent where $f$ is the idealized filling of the metal $d$ band,
$V_{\pi}^2$ is the NO $\pi$-surface coupling matrix element,
$\epsilon_{2\pi^*}$ is the energy of the renormalized 2$\pi$* orbital
relative to the Fermi level, and $\epsilon_d$ is the first moment of
the metal $d$-projected density of states relative to the Fermi level.
The HMN model predicts that larger values of $V_{\pi}^2$ give rise to
greater back-donation, and based on overlap arguments $V_{\pi}^2$ is
expected to increase with the extent of CO-metal coordination.  As
shown in Figure~\ref{fig:TM(111)}, the (111) surface of fcc transition
metals has four high-symmetry adsorption sites, which in order of
increasing coordination are top (T) $<$ bridge (B) $<$ hcp hollow (H) $=$ fcc
hollow (F). Thus, based on the HMN model interpretation of the
CO/metal puzzle, the 3-fold H and F sites will exhibit greater DFT-GGA
overestimation of $E_{\rm chem}^{d \rightarrow 2\pi^*}$ than the
doubly coordinated B site or singly coordinated T site.

Several studies have traced the CO/metal puzzle to its
origins~\cite{Grinberg02p2264,Gil03p71,Kresse03p073401}, and various
schemes have been developed to correct or address
it~\cite{Olsen03p4522,Kresse03p073401,Abild-Pedersen07p1747,Dabo07p11045}. One
approach, a post-DFT extrapolation procedure, was developed by
Grinberg and Rappe in collaboration with one of the co-authors of this
study~\cite{Mason04p161401R}. The extrapolation procedure was based on
the linear relationship between $E_{\rm chem}$ and the CO
singlet-triplet excitation energy $\Delta E_{\rm T \leftarrow S}$, and
was developed for different adsorption sites on (111) and (100)
transition metal surfaces of Pt, Pd, Rh, and Cu.  The linear trends
were determined by changing the design of carbon and oxygen
pseudopotentials sufficiently to achieve variation in the predicted
properties.  The linear trends obtained from the DFT-GGA calculations
were then extrapolated to a $\Delta E_{\rm T \leftarrow S}$ value from
a configuration interaction calculation.  Finally, the corresponding
ordinates were taken to be the corrected values of $E_{\rm
  chem}$. Consistent with the above conceptual argument that the
DFT-GGA error in $E_{\rm chem}^{d\rightarrow 2\pi^*}$ will be larger
for higher NO-metal coordination, the slope $\delta E_{\rm chem}^{\rm
  GGA}$ / $\delta \Delta E_{\rm T \leftarrow S}^{\rm GGA}$ was found
to have the greatest magnitude at hollow sites and the smallest
magnitude at top sites.

The accuracy of DFT-GGA predictions of $E_{\rm chem}$ for NO on
transition metals has been given less consideration than in the
analogous case of CO.  This may be due in part to the fact that CO
preferentially adsorbs at top sites on several late transition metal
surfaces.  As the nature of the CO/metal puzzle is to overestimate
$E_{\rm chem}$ at highly coordinated sites more than at top sites,
this leads to cases of incorrect site preference predictions for CO in
uncorrected DFT-GGA
calculations~\cite{Gil03p71,Lynch00p1,Mitas00p1479,Grossman95p3870}. On
the other hand, NO exhibits an inherent preference to adsorb at highly
coordinated sites on most late transition metal surfaces.  Therefore,
the same overestimation of back-donation in NO/metal systems will not
result in the gross qualitative error of incorrect site preference as
seen in many CO/metal systems.  Instead, overestimation of $E_{\rm
  chem}^{d \rightarrow 2\pi^*}$ in NO/metal systems will only manifest
quantitatively in calculated values of $E_{\rm chem}$, in a
site-specific fashion.  In addition to there being no obvious problem
with DFT-GGA predictions of NO site preference, there are theoretical
arguments for why overestimation of back-donation may not not occur.
For example, in a comprehensive DFT study~\cite{Gajdos06p13}, a detailed
comparison of CO and NO chemisorption is presented, and it is
suggested that the partial occupation of the NO 2$\pi^{*}$ orbital
effectively ``pins'' the energy of that state to the Fermi level,
which prevents overestimation of back-donation.

In the present study, we demonstrate that in most adsorption
geometries, the rehybridization of the NO 2$\pi^{*}$ orbital in the
chemisorbed state can give rise to new bonding and anti-bonding
states.  The NO/metal 2$\pi^{*}$ rehybridization is compared to that
in CO/metal systems, and we interpret the results to support that
DFT-GGA overestimation of $E_{\rm chem}^{d \rightarrow 2\pi^*}$ does
affect predictions for $E_{\rm chem}$ in NO/metal systems.  We
demonstrate a linear relationship between NO $E_{\rm chem}$ and
$\Delta E_{\rm G\leftarrow X}$, the energy of the
G$^{2}\Sigma^{-}\leftarrow \rm{X}^{2}\Pi$ excitation of the free NO
molecule.  We use these findings as the basis for the development of a
post-DFT extrapolation scheme in the style of Mason and
co-workers~\cite{Mason04p161401R}.  Corrected values for $E_{\rm
  chem}$ do not in general affect the predicted NO site preference,
but do quantitatively change the values of $E_{\rm chem}$ by up to
18\%.

\section{\label{sec:level1}Methodology}

Calculations are carried out using an in-house solid state DFT code as
used in previous
studies~\cite{Mason04p161401R,Mason06p3816,Mason08p1963,Bennett06p180102R}
employing the PBE-GGA exchange-correlation functional. Norm-conserving
optimized (RRKJ)~\cite{Rappe90p1227} pseudopotentials with the
designed nonlocal method for metals~\cite{Ramer99p12471} or
Troullier and Martins (TM) style~\cite{Troullier91p1993} norm-conserving
pseudopotentials are employed, generated using the OPIUM
pseudopotential package~\cite{Opium2}. In all DFT calculations, the
Kohn-Sham orbitals are expanded in a plane-wave basis set truncated at
50~Ry.

\begin{figure}
\includegraphics[width=6.0in]{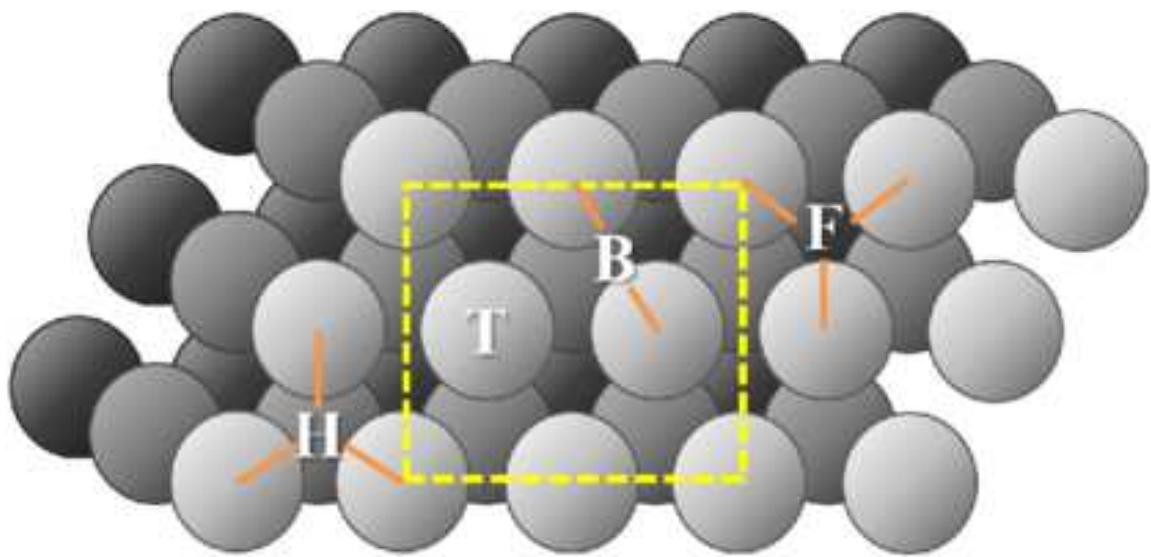}
\caption{{Top view of the (111) fcc transition metal surface with high-symmetry adsorption site labels.  Top (T), bridge (B), hcp (H), fcc (F) sites are labeled.  The $c$(4$\times$2) cell is indicated by dashed lines.}}
\label{fig:TM(111)}
\end{figure}

NO Chemisorption is modeled on the (111) surfaces of Rh, Pd, Ir and
Pt. The theoretical lattice constants of face-centered unit cell for
those metal species are determined to be 3.797, 3.867, 3.840 and
3.954~\AA, respectively, based on polynomial fits and minimization of
primitive bulk cell total energy {\it vs.} volume data points. Models
for the (111) surfaces are generated in $c$(4$\times$2) supercells
(shown in Figure~\ref{fig:TM(111)}) comprised of five atomic layers of
metal and at least 15.6~\AA\ of vacuum separating periodic images in
the surface normal direction. In all surface calculations, the first
Brillouin zone is sampled using a converged 4$\times$4$\times$1 grid
of Monkhorst-Pack $k$-points~\cite{Monkhorst76p5188}. NO chemisorption
is modeled at 1/4 ML coverage in each of the four adsorption sites T,
B, H, and F (Figure~\ref{fig:TM(111)}), with NO bonding perpendicular
to the surface (except as noted) and through N. Geometry optimization
for each structure, allowing for relaxation in the top two metal
layers, is performed until the highest residual force is no larger
than 0.01 eV/\AA.  For the T site, two sets of calculations are
carried out, either imposing linear (T$_{\rm l}$) or allowing for bent
(T$_{\rm b}$) chemisorbed NO geometries.

In order to track changes in values of $E_{\rm chem}$ with variations
in molecular NO electronic structure, calculations of the NO/metal
systems using 4 distinct O pseudopotentials are carried out, with
details of the O pseudopotential designs given in Table~\ref{tab:psp}.

Values for $E_{\rm chem}$ are calculated using

\begin{eqnarray}
E_{\rm chem} =E_{\rm NO/surface} - E_{\rm NO} - E_{\rm surface}.
\label{eqn:Echem}
\end{eqnarray}

\noindent With this definition, chemisorption energies are reported in
eV/NO molecule, and negative values of $E_{\rm chem}$ represent
favorable adsorption on the surface.

Atomic state-by-state projected density of states (PDOS) analysis is
carried out by by projecting atomic valence pseudo-wavefunctions
(radial wavefunction multiplied by real combination of spherical
harmonics) of atoms onto all the Kohn-Sham orbitals.  Band centers are
then calculated as the first moment of each projection.  Here, we are
interested in monitoring how the states of NO change in going from the
free molecule to the chemisorbed state.  To approximate the projection
onto NO $\pi$ and $\sigma$ orbitals, we follow the approach of Zeng and
co-workers~\cite{Zeng10p2459} and take linear combinations of N and O
PDOS:

\begin{eqnarray}
\pi = {\rm N}_{p_{x}} +  {\rm N}_{p_{y}} +  {\rm O}_{p_{x}} +  {\rm N}_{p_{y}},
\label{eqn:pi}
\end{eqnarray}

\begin{eqnarray}
\sigma = {\rm N}_{s} +  {\rm N}_{p_{z}} +  {\rm O}_{s} +  {\rm N}_{p_{z}}.
\label{eqn:sigma}
\end{eqnarray}

\section{\label{sec:level1} Results and Discussion}    

In our calculation using the O pseudopotential labeled PSP 1 in
Table~\ref{tab:psp}, the N-O bond length, $d_{\rm N-O}$ of the free NO
molecule is 1.148~\AA.  This is in good agreement with the
experimental value of 1.154~\AA, though notably shorter than other
DFT-GGA calculated results between 1.16~\AA and
1.18~\AA~\cite{Zhanpeisov06p801,Koper00p4392,Gajdos06p13,Getman07p389,Aizawa02p394,Loffreda98p6447,Loffreda98p15,Zeng09p205413,Rempel09p20623,Ge98p15}. We
attribute the difference in $d_{\rm N-O}$ to pseudopotential style, as
the above referenced DFT-GGA values are all obtained using ultrasoft
pseudoptoentials, while our study employs norm-conserving optimized
pseudopotentials.  The values for $d_{\rm N-O}$ and other aspects of
the optimized NO/metal geometries using O PSP 1 are presented in
Table~\ref{tab:Geometry}. For each metal surface and site, $d_{\rm
  N-O}$ is up to 3.4\% shorter than literature values for studies
employing ultrasoft
pseduoptoentials~\cite{Zeng09p205413,Rempel09p20623,Popa06p245408,Herron12p1670,Krekelberg04p987,Burch02p2902,Tang05p17630},
supporting systematic disagreement in $d_{\rm N-O}$ between the two
pseudoptoential approximations. For all metals, as expected, $d_{\rm
  N-O}$ increases with increasing NO-metal coordination. Chemically
speaking, the bond order of chemisorbed NO decreases as $d_{\rm N-O}$
increases.

\begin{figure}
\includegraphics[width=6.0in]{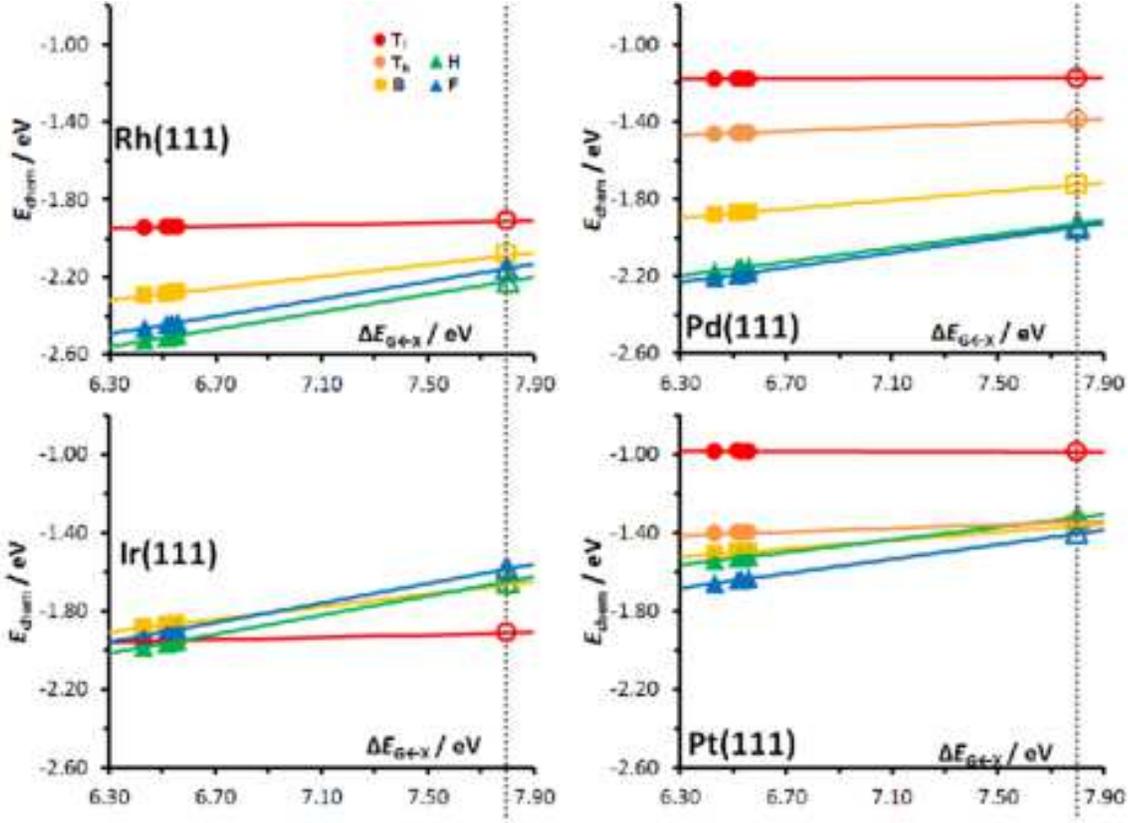}
\caption{{(Color online) Plots of $E_{\rm chem}$ $vs.$ $\Delta E_{\rm G\leftarrow X}$ for calculations using the four sets of O pseudopotentials (in Table~\ref{tab:psp}) for 1/4 ML NO at T$_{\rm l}$ (red circles), T$_{\rm b}$ (orange circles), B (yellow squares), H (green triangles), and F (blue triangles) sites on Rh, Pd, Ir, and Pt(111) surfaces. Values of $E_{\rm chem}$ are indicated by filled symbols, while the corresponding values of $E_{\rm chem}^{\rm corr}$ are shown by open symbols.  The experimental value for $\Delta E_{\rm G\leftarrow X}$ 62913.0 cm$^{\rm -1}$ (7.80~eV) is indicated in each plot by a vertical dotted line.}}
\label{fig:Trends}
\end{figure}

As shown in Figure~\ref{fig:Trends}, calculations for $E_{\rm chem}$
using the four distinct O pseudopotentials exhibit a linear
relationship between $E_{\rm chem}$ and $\Delta E_{\rm G\leftarrow
  X}$, similar to the linear relationships previously reported between
CO $E_{\rm chem}$ and $\Delta E_{\rm T \leftarrow
  S}$~\cite{Mason04p161401R} or CO $E_{\rm chem}$ and the energy of
the CO 2$\pi^{*}$ orbital~\cite{Kresse03p073401}. The reason for the
correlation between NO $E_{\rm chem}$ and $\Delta E_{\rm G\leftarrow X}$
is considered in terms of the charge transfer in the chemisorption and
molecular excitation processes.  The G$^{2}\Sigma^{-}\leftarrow
\rm{X}^{2}\Pi$ excitation involves electron transfer from the NO
5$\sigma$ orbital to 2$\pi^{*}$ orbital, while back-donation in
chemisorption involves charge transfer from the metal $d$-band into
the 2$\pi^{*}$-derived bonding states.

In a style similar to Mason {\it et al.}~\cite{Mason04p161401R}, we
extrapolate corrected values of the chemisorption energy $E_{\rm
  chem}^{\rm corr}$, as shown in Figure~\ref{fig:Trends}. The
extrapolation process to calculate $E_{\rm chem}^{\rm corr}$ is
summarized as

\begin{eqnarray}
E_{\rm chem}^{\rm corr} =E_{\rm chem} + (\Delta E_{\rm G\leftarrow X}^{\rm exp} - \Delta E_{\rm G\leftarrow X}^{\rm GGA}) \frac {\delta E_{\rm chem}} {\delta \Delta E_{\rm G\leftarrow X}^{\rm GGA}},
\label{eqn:Corr}
\end{eqnarray}

\noindent where $\Delta E_{\rm G\leftarrow X}^{\rm exp}$ and $\Delta
E_{\rm G\leftarrow X}^{\rm GGA}$ are the experimental and DFT-GGA
values for the NO G$^{2}\Sigma^{-}\leftarrow \rm{X}^{2}\Pi$
excitation, respectively.  Her, we use a reported
experimental value for $\Delta E_{\rm G\leftarrow X}$ 62913.0 cm$^{\rm
  -1}$ (7.80~eV) as the calibration~\cite{HerzbergCDM}.  $\delta E_{\rm chem}$ / $\delta \Delta
E_{\rm G\leftarrow X}^{\rm GGA}$ indicates the so-called correction slope of the
linear trend. All the $E_{\rm chem}^{\rm corr}$ results are given in
Table~\ref{tab:Echem}, alongside the uncorrected $E_{\rm chem}$ values
obtained using O PSP 1.

An overall picture picture presented by Figure~\ref{fig:Trends} and
Table~\ref{tab:Echem} is that values of $E_{\rm chem}^{\rm corr}$ are
smaller in magnitude than uncorrected $E_{\rm chem}$, and the
correction slope has a significant dependence on the adsorption
site. The trends in $\delta E_{\rm chem}$ / $\delta \Delta E_{\rm
  G\leftarrow X}^{\rm GGA}$ by site can be rationalized in terms of
NO-metal coordination and how the orbitals of adsorbate and substrate
overlap. As discussed in the
literature~\cite{Gajdos06p13,Zeng10p2459}, for T$_{\rm l}$ adsorption,
the greatest contribution to the NO-metal overlap is from the
5$\sigma$ orbital of NO molecule (comprised of N and O atomic $s$ and
$p_{\rm z}$ orbitals), and the $d_{\rm z^{\rm 2}}$-derived states of
the chemisorbing surface atom. When there are two or three surface
atoms coordinated with NO, such as at B, H, or F sites, the metal
atom-N-O angle is not linear, and this change in chemisorption
geometry results in different dominant overlap interactions.
Specifically, in B, H, or F geometries, the NO-metal overlap is
dominated by interaction of the NO 2$\pi^{*}$ orbital (comprised of N
and O atomic $p_{\rm x}$ and $p_{\rm y}$ orbitals), and the $d_{\rm
  xz}$- and $d_{\rm yz}$-derived states of the chemisorbing surface
atoms.  Thus, as NO-metal coordination increases, the overlap between
NO 2$\pi^{*}$ and the surface also increases, which causes the
chemisorption bond to have more back-donation character.  This leads
to overestimation of $E_{\rm chem}^{d \rightarrow 2\pi^*}$, the
relative extent of which can be tracked by the relative values of
$\delta E_{\rm chem}$ / $\delta \Delta E_{\rm G\leftarrow X}^{\rm
  GGA}$ for different sites.

As foreshadowed, the result of extrapolating to $E_{\rm chem}^{\rm
  corr}$ does not impact the predicted NO site preference in most cases,
but does affect the magnitude of $E_{\rm chem}$ by up to 18\%.  The
most favorable site before and after the correction is indicated in
Table~\ref{tab:Echem}.  The one instance of altered site preference is
for Ir(111), and this can be explained by the very close spacing of
uncorrected $E_{\rm chem}$ values on that surface.  Qualitatively, our
results for site preference are in agreement with previous theoretical
results~\cite{Zhanpeisov06p801,Koper00p4392,Gajdos06p13,Getman07p389,Aizawa02p394,Loffreda98p6447,Loffreda98p15,Zeng09p205413,Rempel09p20623,Popa06p245408,Herron12p1670,Krekelberg04p987,Burch02p2902,Tang05p17630,Ge98p15,Ford05p159,
  Mavrikakis02p6737,Honkala01p72,Honkala01p5942,Hansen02p1,Zeng10p2459,Deshlahra12p8408,Zeng10p085408}.
Quantitatively, the values of $E_{\rm chem}^{\rm corr}$ reported here
may be more reliable, and in particular the relative values of $E_{\rm
  chem}^{\rm corr}$ at different sites on the same metal may be
important for interpreting how chemisorption evolves with coverage.
For example, in the uncorrected values, the range of $E_{\rm chem}$
for different sites on Rh(111) is 0.584~eV, while the range in $E_{\rm
  chem}^{\rm corr}$ is 0.310.  Thus the corrected values predict a
much smoother potential energy surface than the uncorrected values.

\begin{figure}
\includegraphics[width=6.0in]{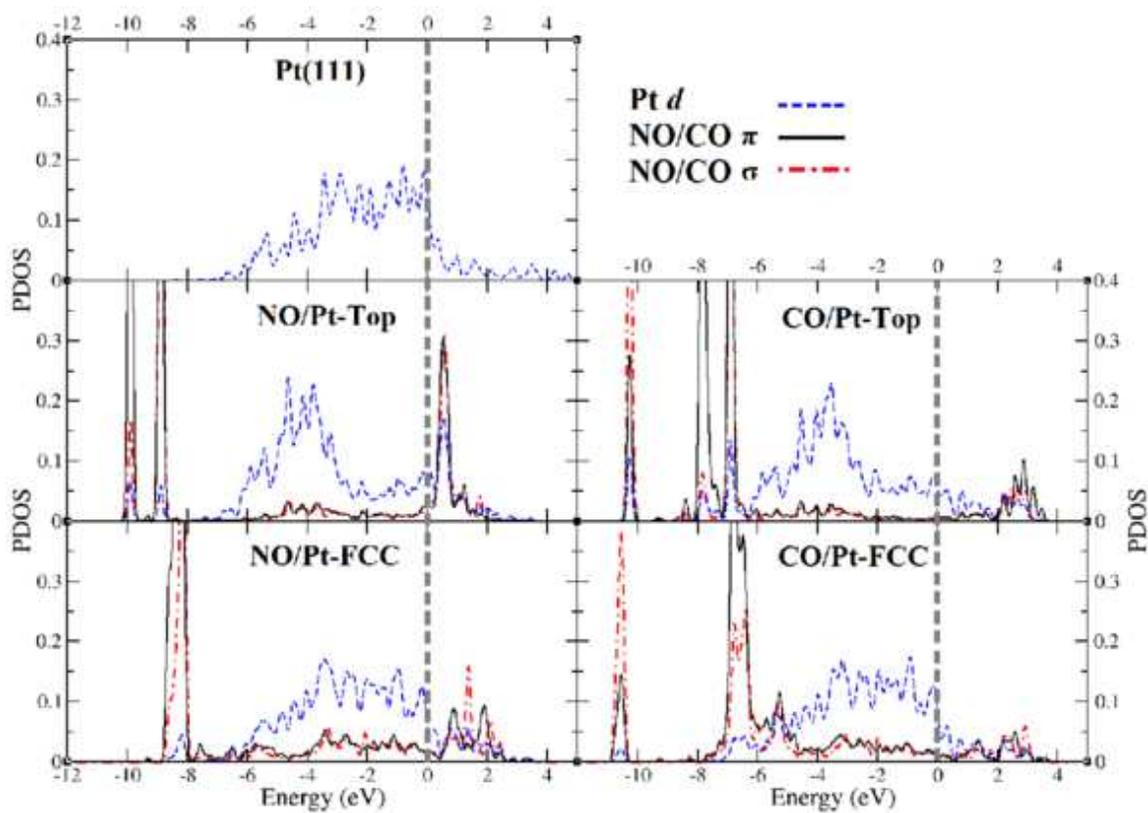}
\caption{{(Color online) Projected density of states (PDOS) for bare Pt(111) surface and NO and CO adsorption complexes at T$_{\rm l}$ and F sites. The $\pi$ and $\sigma$ projections are calculated according to the Equations~\ref{eqn:pi} and~\ref{eqn:sigma}, respectively. The Fermi level is referenced to zero and is indicated by vertical dashed lines in each plot.}}
\label{fig:PDOS}
\end{figure}

The basis of our assertion that there is DFT-GGA overestimation of
$E_{\rm chem}^{d \rightarrow 2\pi^*}$ relies on the NO 2$\pi^{*}$
orbital rehybridizing with the surface $d$-band to form new bonding
and anti-bonding states in the NO/metal system, similar to as seen in
CO chemisorption.  This is at odds with the conclusion reported by
Gajdos and co-workers~\cite{Gajdos06p13}, which alternatively states
that the NO 2$\pi^{*}$-derived states in NO/metal systems span a
single, narrow energy range close to the Fermi level.  In order to
clarify the bonding interactions in NO/metal systems, we carry out
comparative density of states analysis.  In Figure~\ref{fig:PDOS}, the
projections onto the Pt $d$, NO $\pi$, and NO $\sigma$ states is shown
for NO/Pt(111) T$_{\rm l}$ and F sites. Consistent with the
interpretation of Gajdos {\it et al.}, inspection of the PDOS for the
T$_{\rm l}$ site shows that the 2$\pi^{*}$ state, which we assign as
the narrow and intense peak in the NO $\pi$ PDOS just above the Fermi
level, does not significantly hybridize with the Pt $d$-band. However,
in the PDOS of the F configuration, the sharp peak in the NO $\pi$
projection near the Fermi level is diminished, and new broad intensity
appears at bonding energies of 0-4~eV below the Fermi level.  Thus the
PDOS provides evidence of charge transfer from the surface into NO
2$\pi^{*}$-derived states at the F site, which also gives rise to a
relatively large correction slope value of 0.19.  On the other hand,
the absence of PDOS evidence for 2$\pi^{*}$-Pt hybridization in the
T$_{\rm l}$ geometry is reflected in the negligible value of $\delta
E_{\rm chem}$ / $\delta \Delta E_{\rm G\leftarrow X}$ for the T$_{\rm
  l}$ site. 

For comparison, the PDOS of CO/Pt(111) T$_{\rm l}$ and F are presented
in Figure~\ref{fig:PDOS}. A key distinction between the PDOS of
NO/Pt(111) and CO/Pt(111) is noted for the T$_{\rm l}$ configurations:
In CO/Pt(111) T$_{\rm l}$, the CO 2$\pi^{*}$ state does hybridize with
the surface, while this behavior is absent in the NO/Pt(111) T$_{\rm
  l}$. The fact that hybridization of the molecular 2$\pi^{*}$ states
occurs at the F site but does not occur at the T$_{\rm l}$ site in
NO/Pt(111) may be why it has been previously thought that NO/metal
systems are not subject to the same overestimation of $E_{\rm chem}^{d
  \rightarrow 2\pi^*}$ as confirmed in CO/metal systems. The PDOS for
NO/Pt(111) B and H sites, not shown, show behavior qualitatively
similar to NO/Pt(111) H.

The chemisorption-induced charge transfer is further investigated by
studying the changes in the occupation of NO and CO $\pi$ and $\sigma$
states between the gas-phase and chemisorbed systems.  Using an
analysis similar to as done by Zeng and co-workers,~\cite{Zeng10p2459}
fractional fillings of the chemisorbed NO $\sigma$ and $\pi$
projections are calculated and compared to the ideal molecular values.
The changes in $\pi$ occupation, $\Delta \pi$, and in $\sigma$
occupation, $\Delta \sigma$, are reported in Table~\ref{tab:Charge},
along with the sum net change in both occupations.  For all of the NO
and CO chemisorption systems, the positive values of $\Delta \pi$
reflect increased occupation of the $\pi$ states upon chemisorption,
while the negative values of $\Delta \sigma$ reflect decreased
occupation of the $\sigma$ states.  The magnitude of ($\Delta \pi +
\Delta \sigma$) increases with NO-metal coordination. The trends in
$\Delta \pi$ and $\Delta \sigma$ for NO and CO support that
simultaneous donation and back-donation contribute to the
chemisorption process of both molecules, in agreement with other
theoretical descriptions of NO (as in
References~\cite{Getman07p389,Zeng09p205413,Zeng10p2459}) and CO (as
in References~\cite{Koper00p4392,Hammer96p2141})
chemisorption. Comparing the NO/Pt(11) and CO/Pt(111) systems, the
larger $\Delta \pi$ values of the latter may be attributed to the
empty occupation of the molecular CO 2$\pi^{*}$ orbital.  The
comparative values of $\Delta \pi$ are also consistent with the
relative magnitude of the correction slopes in NO and CO systems.  For
example, as previously reported for CO, the correction slopes for Rh,
Pd and Pt(111) surfaces can be up to 0.26 for top site and 0.56 for
hollow site~\cite{Mason04p161401R}, while in this work that the
correction slopes are near 0 for T$_{\rm l}$ sites and up to 0.25 for
H and F sites (Table~\ref{tab:Echem}).

\begin{figure}
\includegraphics[width=6.0in]{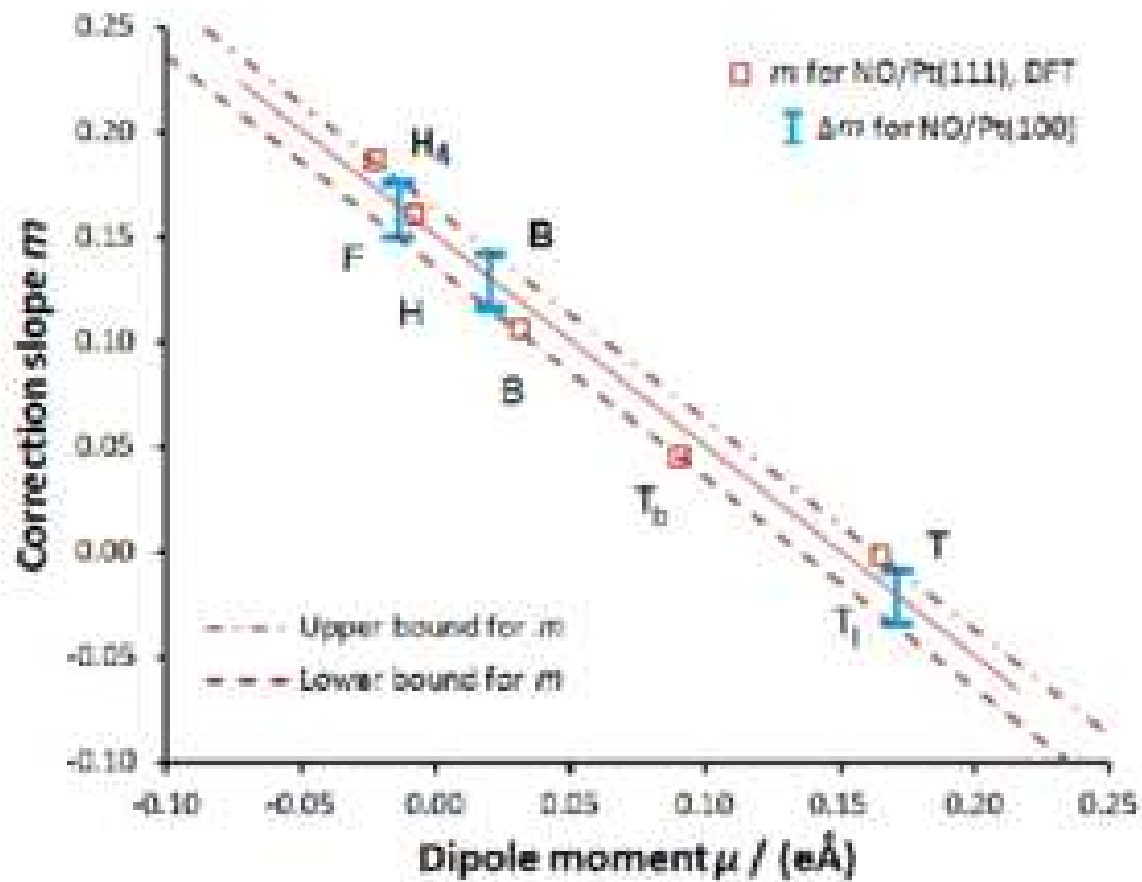}
\caption{{(Color online) Plot of the correlation between $E_{\rm chem}$ correction slopes and the dipole moment $\mu$ calculated using O PSP 1 for the five sites on Pt(111) surface. The NO/Pt(111) DFT data are shown with red crosses. The approximate correction slope $m$ is bound by upper and lower limits shown as red dashed lines. The range for the approximate Pt(100) correction slope values, $\Delta m$ are shown in blue for three different adsorption sites.}}
\label{fig:slope/dipole}
\end{figure}

The presented correction scheme requires calculations using multiple
pseudopotentials to determine the values of $\delta E_{\rm chem}$ /
$\delta \Delta E_{\rm G\leftarrow X}$.  In the CO extrapolation
procedure, it was noted that the correction slope values correlated
with the CO vibrational stretch frequency values for each site, with
little dependence on the metal identity.  While this afforded an
approximate correction scheme for CO/metal systems, the overlapping
ranges for the NO vibrational stretch frequency values at different
sites as discussed in References~\cite{Gajdos06p13} and~\cite{Greeley02p319} precludes such a short-cut for approximating
$\delta E_{\rm chem}$ / $\delta \Delta E_{\rm G\leftarrow X}$.
However, we note that there is a correlation between the correction
slope and calculated dipole moment, $\mu$.  Here we
define the electron distribution of N$^{\delta -}$-O$^{\delta +}$ as
positive dipole moment, and N$^{\delta +}$-O$^{\delta -}$ as negative
dipole moment. As recently reported and interpreted by Deshlahra and
co-workers~\cite{Deshlahra12p8408}, the sign and magnitude of the
dipole moment of values in NO/metal exhibits site-specific trends, with T
sites having $\mu$ $>$ 0, and H sites having $\mu$ $<$ 0.  The Pt(111)
correction slopes for the five (111) adsorption sites and associated
values of $\mu$ are given in Table~\ref{tab:dipolecorrection}.  As shown in
Figure~\ref{fig:slope/dipole}, our calculated values of $\mu$
demonstrate that the sign of $\mu$ changes from positive to negative
as the NO-metal coordination increases.  Simultaneously, the
correction slopes grow grow larger, consistent with the greater extent
of back-donation at highly coordinated adsorption sites.

\begin{figure}
\includegraphics[width=6.0in]{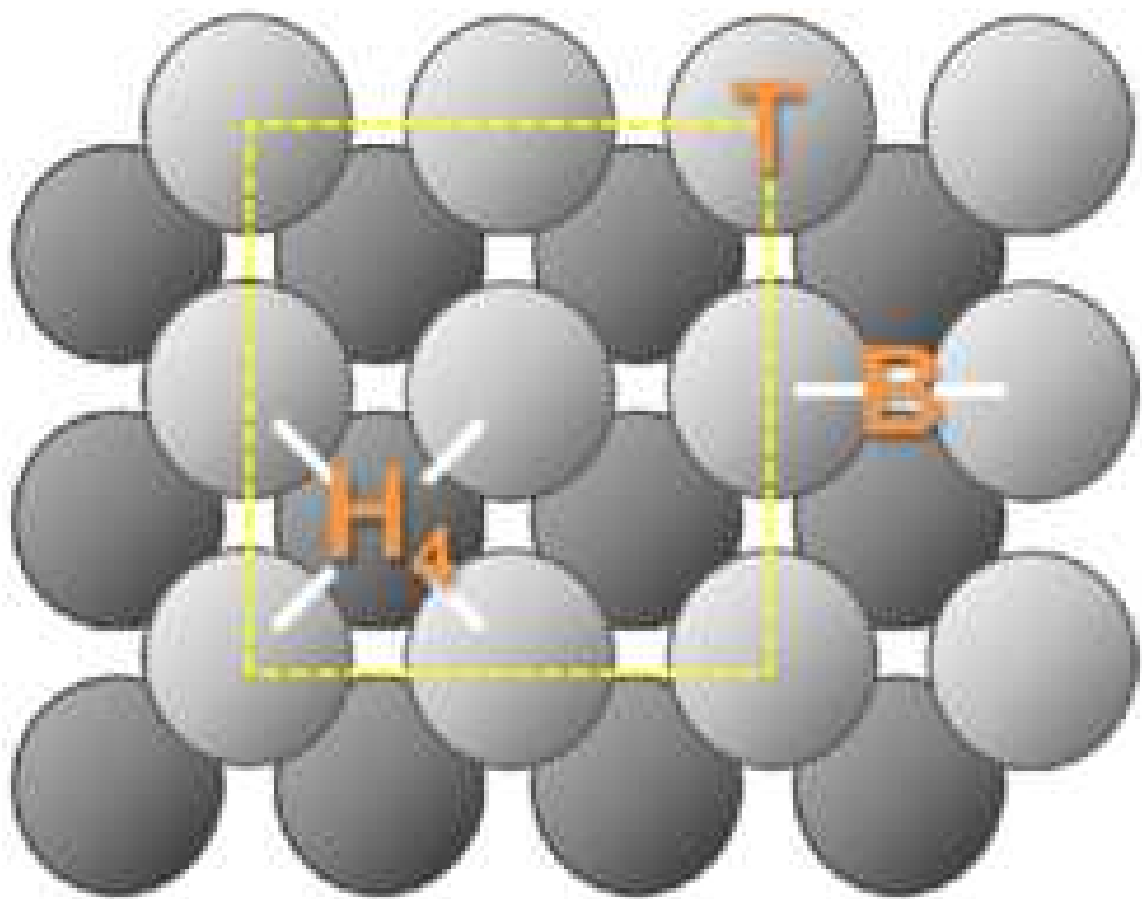}
\caption{{Top view of the Pt(100) surface.  Top (T), bridge (B), and four-fold hollow H$_{4}$ adsorption sites are indicated.  The employed surface cell is shown by dashed lines.}}
\label{fig:Pt(100)}
\end{figure}

The relationship between $\mu$ and $\delta E_{\rm chem}$ / $\delta
\Delta E_{\rm G\leftarrow X}$ enables an approximate scheme for
determining correction slope.  In this way, $E_{\rm chem}^{\rm corr}$
can be extrapolated using a smaller set of DFT calculations.  To test
the $\mu$-based extrapolation scheme, we model the NO/Pt(100)
chemisorption system and determine $E_{\rm chem}^{\rm corr}$ in two
ways, first using the full extrapolation procedure and again by using
the $\mu$-based correction detailed below.  Chemisorption energies
using the $\mu$-based correction are denoted $E_{\rm chem}^{\rm corr, dip}$.  This comparison is done for three adsorption sites
on Pt(100) surfaces shown in Figure~\ref{fig:Pt(100)}: top (T), bridge
(B), and the four-fold hollow site H$_{4}$.

\begin{figure}
\includegraphics[width=6.0in]{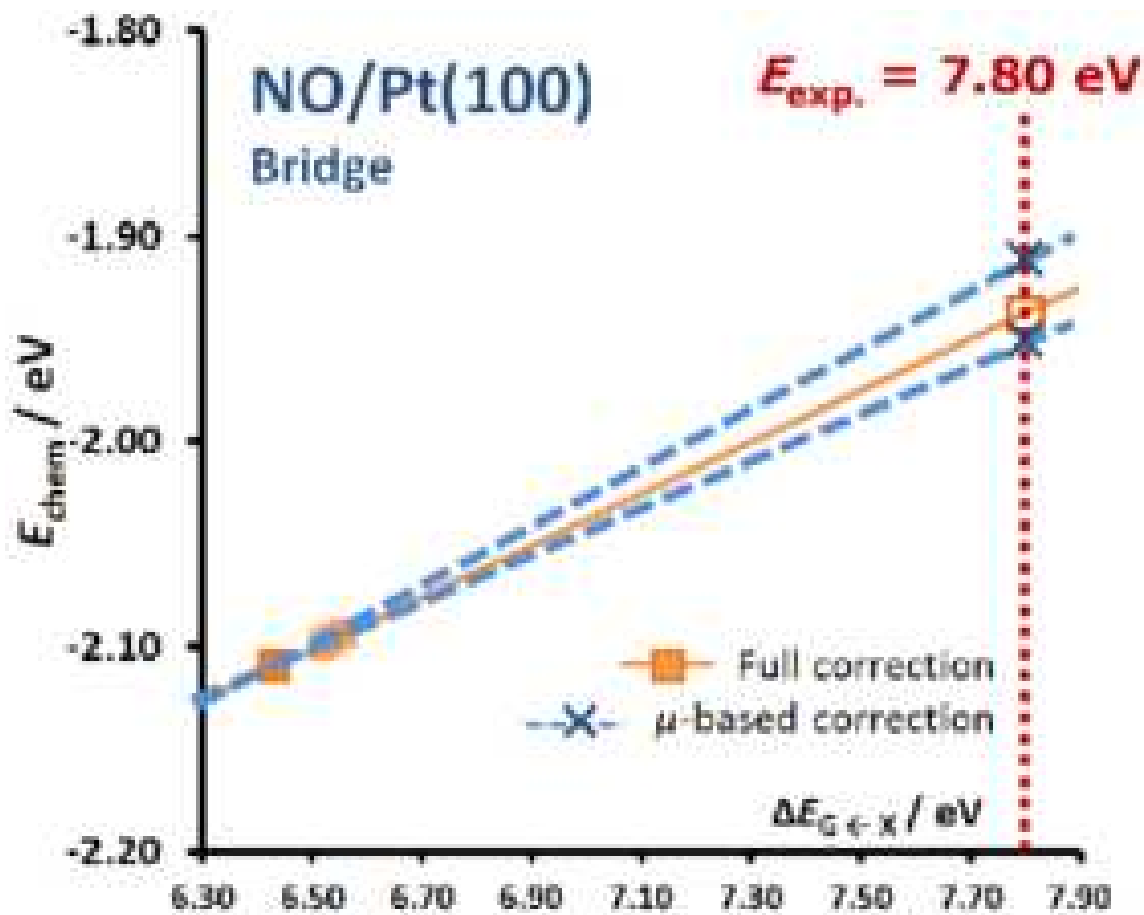}
\caption{{(Color online) Summary of the $\mu$-based correction method for NO/Pt(100) B site. The two blue dashed lines indicate the trends in $E_{\rm chem}$ $vs.$ $\Delta E_{\rm G\leftarrow X}$ using the upper and lower bound correction slope values from Figure~\ref{fig:slope/dipole}. The $\mu$-based correction values for  $E_{\rm chem}^{\rm corr, dip}$ are shown in blue X's.  The uncorrected value of $E_{\rm chem}$ and the corresponding value of $E_{\rm chem}^{\rm corr}$ using the full correction scheme are shown in filled and empty orange squares, respectively, connected by a linear fit shown in orange.  In all cases, the corrected chemisorption energy value is determined by extrapolating the linear trend to the $\Delta E_{\rm G\leftarrow X}$ value of 7.80~eV.}}
\label{fig:dipolecorrection}
\end{figure}

To carry out the dipole-based correction method, we first calculate
$E_{\rm chem}$ and the corresponding dipole moment for NO at the three sites
on Pt(100) using O PSP 1. The NO/Pt(100) values of $\mu$ are used along
with the NO/Pt(111) trend between $\mu$ and $\delta E_{\rm chem}$ /
$\delta \Delta E_{\rm G\leftarrow X}$ to approximate the NO/Pt(100)
correction slope for each site.  Once the value of $\delta E_{\rm
  chem}$ / $\delta \Delta E_{\rm G\leftarrow X}$ is approximated, the
correction method proceeds in the same fashion as detailed for the
full method, using an experimental value for $\Delta E_{\rm
  G\leftarrow X}$ to arrive at values of $E_{\rm chem}^{\rm corr, dip}$.
We suggest a range of NO/Pt(100) correction slope values based on the
dipole correlation, and in Figure~\ref{fig:dipolecorrection} the
resulting upper and lower bound $E_{\rm chem}^{\rm corr}$ values are
shown. Before correction, the $E_{\rm chem}$ value for the Pt(100)
bridge site (reported as the preferred geometry~\cite{Ge04p1551,Mei04p361,Mei10p10364,Ranea06p2663}) is -2.109
eV; after correction, the range of $E_{\rm chem}^{\rm corr, dip}$ is
-1.913 eV to -1.951 eV. As a comparison, the full pseudopotential
correction method is also performed and plotted in
Figure~\ref{fig:dipolecorrection}. The result of -1.937 eV obtained
from the full correction scheme is within the bounds of the
dipole-based approximate correction method, supporting the validity
and utility of the $\mu$-based correction.  Results for the remaining
NO/Pt(100) sites using both the full and $\mu$-based correction method
are given in Table~\ref{tab:dipolecorrection}.

\section{\label{sec:level1} Conclusions}    

NO chemisorption on Rh, Pd, Ir, and Pt(111) surfaces is studied to
assess if over-estimation of back-donation leads to systematic errors
in DFT-GGA predictions for $E_{\rm chem}$. The electronic structure of
NO/metal and CO/metal systems shows that, with the exception of linear
top site adsorption, the NO 2$\pi^{*}$ orbital does rehybridize with
the surface $d$-band to form new bonding and anti-bonding states,
similar to what is seen in the analogous CO/metal systems.  Linear
trends of NO $E_{\rm chem}$ {\it vs.} $\Delta E_{\rm G\leftarrow X}$
are obtained by calculations utilizing different O pseudopotentials,
and a post-DFT correction scheme for NO $E_{\rm chem}$ is constructed
after the style of a CO chemisorption energy extrapolation
procedure~\cite{Mason04p161401R}.  In the case of NO, the $E_{\rm
  chem}$ correction does not impact the predicted site preference.
However, as the impact of the correction varies with adsorption site,
the relative values of $E_{\rm chem}$ for different sites on a given
metal are affected.  In most cases, the corrected values predict smoother
NO/metal potential energy surfaces than the uncorrected values. The
implication of the correction could be meaningful in understanding the
evolution of NO chemisorption with coverage, NO diffusion on surfaces,
or for accurately predicting the relative magnitude of $E_{\rm chem}$
in NO/metal systems.

\section{\label{sec:level1} Acknowledgments}    

S.E.M. thanks the University of Iowa College of Liberal Arts and
Sciences and the Iowa Center for Research by Undergraduate for funding
this work.  We thank Prof. Andrew M. Rappe and his research group for
use of an in-house DFT code and for the online availability of RRKJ
transition metal pseudopotential design details.  Daniel J. Gillette
is acknowledged for carrying out geometry optimization calculations
contributing to this project.

\clearpage

\begin{table}
\caption{Details of oxygen pseudopotential design. Core radii $r_{c}$
  for the O $s$ and $p$ are in $a_o$.  All pseudopotentials were
  created using 50~Ry cutoff energy, and from the $s^2p^4$ reference
  configuration.  For each pseudopotential, the norm-conserving
  pseudopotential type (RRKJ~\cite{Rappe90p1227} or
  TM\cite{Troullier91p1993}) and associated values of $\Delta E_{\rm
    G\leftarrow X}^{\rm GGA}$ in eV are given.}
\begin{tabular}{cccccc}
&\multicolumn{1}{c}{Type}
&\multicolumn{1}{c}{$r_{c^s}$,$r_{c^p}$}
&\multicolumn{1}{c}{$\Delta E_{\rm G\leftarrow X}^{\rm GGA}$/eV}\\
\hline
\hline
PSP 1  & TM    & 1.34, 1.53  & 6.432 \\
PSP 2  & RRKJ  & 1.32, 1.44  & 6.519 \\
PSP 3  & RRKJ  & 1.34, 1.46  & 6.534 \\
PSP 4  & RRKJ  & 1.34, 1.53  & 6.558 \\
\hline
\label{tab:psp}
\end{tabular}
\end{table}

\begin{table}
\caption{Details of the optimized geometries of NO chemisorption
  obtained using O PSP 1. Reported values include the bond length
  between N and O, $d_{\rm N-O}$, the average distance between N and
  coordinated surface metal atom(s), $d_{\rm N-TM}$, and the vertical
  distance between N and the average height of the topmost metal
  surface layer, $h_{\rm N-TM(111)}$.}
\begin{tabular}{rccccc}
&\multicolumn{1}{c}{Site}
&\multicolumn{1}{c}{$d_{\rm N-O}$ /~\AA}
&\multicolumn{1}{c}{$d_{\rm N-TM}$ /~\AA}
&\multicolumn{1}{c}{$h_{\rm N-TM(111)}$ /~\AA} \\
\hline
\hline
Rh(111)& T$_{\rm l}$  & 1.154  & 1.744  & 1.972 \\
       & B          & 1.183  & 1.931  & 1.464 \\
       & H          & 1.200  & 2.001  & 1.282 \\
       & F          & 1.200  & 2.008  & 1.295 \\
\hline            
Pd(111)& T$_{\rm l}$  & 1.152  & 1.806  & 1.949 \\
       & T$_{\rm b}$  & 1.161  & 1.843  & 2.032 \\
       & B          & 1.177  & 1.927  & 1.426 \\
       & H          & 1.191  & 1.991  & 1.249 \\
       & F          & 1.194  & 1.987  & 1.227 \\
\hline            
Ir(111)& T$_{\rm l}$  & 1.153  & 1.757  & 1.991 \\
       & B          & 1.183  & 1.996  & 1.535 \\
       & H          & 1.203  & 2.067  & 1.355 \\
       & F          & 1.203  & 2.078  & 1.377 \\
\hline            
Pt(111)& T$_{\rm l}$  & 1.150  & 1.836  & 2.012 \\
       & T$_{\rm b}$  & 1.158  & 1.940  & 2.131 \\
       & B          & 1.174  & 2.003  & 1.518 \\
       & H          & 1.187  & 2.098  & 1.359 \\
       & F          & 1.192  & 2.087  & 1.321 \\
\hline
\label{tab:Geometry}
\end{tabular}
\end{table}

\clearpage

\begin{table}
\caption{Results of linear regression of chemisorption energy {\it vs.}  NO  $\Delta E_{\rm G\leftarrow X}$, dipole moment $\mu$, and $\mu$-based correction details for NO/Pt(100). The DFT-GGA values for $E_{\rm chem}$ are given, along with the corrected energies obtained by extrapolation, $E_{\rm chem}^{\rm corr}$. For convenience, the correction change in $E_{\rm chem}$ (obtained from O PSP 1), $\Delta$, is listed.  For each surface, the predicted preferred site is indicated by $E_{\rm chem}$ values in boldface.}
\begin{tabular}{rccccc}
&\multicolumn{1}{c}{Site}
&\multicolumn{1}{c}{Slope}
&\multicolumn{1}{c}{$E_{\rm chem}$/eV}
&\multicolumn{1}{c}{$E_{\rm chem}^{\rm corr}$/eV}
&\multicolumn{1}{c}{$\Delta$/eV} \\
\hline
\hline
Rh(111)& T$_{\rm l}$  & 0.024  & -1.945  & -1.912  & 0.033 \\ 
       & B          & 0.153  & -2.298  & -2.089  & 0.209 \\ 
       & H          & 0.224  & {\bf -2.529}  & {\bf -2.222}  & 0.307 \\ 
       & F          & 0.225  & -2.463  & -2.156  & 0.307 \\
\hline
Pd(111)& T$_{\rm l}$  & 0.003  & -1.179  & -1.175  & 0.004 \\ 
       & T$_{\rm b}$  & 0.051  & -1.463  & -1.393  & 0.070 \\ 
       & B          & 0.114  & -1.883  & -1.728  & 0.155 \\ 
       & H          & 0.178  & -2.173  & -1.930  & 0.243 \\ 
       & F          & 0.190  & {\bf -2.207}  & {\bf -1.947}  & 0.260 \\ 
\hline
Ir(111)& T$_{\rm l}$  & 0.030  & -1.955  & {\bf -1.914}  & 0.041 \\ 
       & B          & 0.160  & -1.887  & -1.668  & 0.219 \\
       & H          & 0.245  & {\bf -1.986}  & -1.651  & 0.335 \\
       & F          & 0.249  & -1.928  & -1.587  & 0.341 \\
\hline
Pt(111)& T$_{\rm l}$  &-0.002  & -0.984  & -0.986  &-0.002 \\ 
       & T$_{\rm b}$  & 0.046  & -1.408  & -1.345  & 0.063 \\
       & B          & 0.106  & -1.509  & -1.363  & 0.146 \\
       & H          & 0.161  & -1.543  & -1.323  & 0.220 \\
       & F          & 0.187  & {\bf -1.659}  & {\bf -1.403}  & 0.256 \\
\hline
\label{tab:Echem}
\end{tabular}
\end{table}

\clearpage

\begin{table}
\caption{Calculated chemisorption-induced changes in the NO $\pi$ and $\sigma$ fillings, $\Delta \pi$ and $\Delta \sigma$, in NO/metal configurations relative to the respective gas-phase molecules.}
\begin{tabular}{rccccc}
&\multicolumn{1}{c}{Site}
&\multicolumn{1}{c}{$\Delta \pi$}
&\multicolumn{1}{c}{$\Delta \sigma$}
&\multicolumn{1}{c}{$\Delta \pi + \Delta \sigma$} \\
\hline
\hline
NO/Rh(111)& T$_{\rm l}$  & 1.04  & -0.93  & 0.11 \\
          & B          & 1.26  & -0.99  & 0.27 \\
          & H          & 1.16  & -0.77  & 0.39 \\
          & F          & 1.17  & -0.77  & 0.39 \\
\hline
NO/Pd(111)& T$_{\rm l}$  & 0.98  & -1.00  & -0.01 \\
          & T$_{\rm b}$  & 0.63  & -0.43  & 0.20 \\
          & B          & 1.23  & -0.99  & 0.23 \\ 
          & H          & 1.15  & -0.78  & 0.36 \\ 
          & F          & 1.16  & -0.77  & 0.38 \\
\hline
NO/Ir(111)& T$_{\rm l}$  & 1.06  & -0.91  & 0.15 \\
          & B          & 1.37  & -1.05  & 0.32 \\
          & H          & 1.20  & -0.72  & 0.48 \\
          & F          & 1.22  & -0.71  & 0.50 \\
\hline
NO/Pt(111)& T$_{\rm l}$  & 1.00  & -0.96  & 0.05 \\
          & T$_{\rm b}$  & 0.63  & -0.26  & 0.36 \\
          & B          & 1.40  & -1.02  & 0.38 \\
          & H          & 1.23  & -0.69  & 0.54 \\
          & F          & 1.27  & -0.66  & 0.61 \\
\hline
\hline
CO/Pt(111)& T$_{\rm l}$  & 1.57  & -0.49  & 1.08 \\
          & T$_{\rm b}$  & 1.57  & -0.49  & 1.08 \\
          & B          & 1.69  & -0.37  & 1.33 \\
          & H          & 1.69  & -0.46  & 1.23 \\
          & F          & 1.68  & -0.47  & 1.21 \\
\hline
\label{tab:Charge}
\end{tabular}
\end{table}

\begin{table}
\caption{Results of linear regression of chemisorption energy {\it vs.} NO $\Delta E_{\rm G\leftarrow X}$. The DFT-GGA values for $E_{\rm chem}$ and corresponding dipole moments $\mu$ (obtained from O PSP 1) are given, along with the suggested range of correction slope values $\Delta m$.  The resulting range of $\mu$-based $E_{\rm chem}^{\rm corr}$ and changes relative to uncorrected energies $\Delta^{\rm dip}$ are listed. For comparison, the $E_{\rm chem}^{\rm corr}$ obtained from the full correction method are also given. For each surface, the predicted preferred site is indicated by $E_{\rm chem}$ values in boldface.}
\begin{tabular}{rcccccccc}
&\multicolumn{1}{c}{Site}
&\multicolumn{1}{c}{$\mu$/e~\AA}
&\multicolumn{1}{c}{$E_{\rm chem}$/eV}
&\multicolumn{1}{c}{$\Delta m$}
&\multicolumn{1}{c}{Range of $E_{\rm chem}^{\rm corr, dip}$/eV}
&\multicolumn{1}{c}{$\Delta^{\rm dip}$/eV}
&\multicolumn{1}{c}{$E_{\rm chem}^{\rm corr}$/eV} \\
\hline
\hline
Pt(100)& T$_{\rm l}$  & 0.171  & -1.420  & [-0.035, -0.008]  & [-1.468, -1.431]  & [-0.048, -0.011]  & -1.431 \\ 
       & B          & 0.021  & {\bf -2.109}  & [0.116, 0.143]    & {\bf [-1.951, -1.914]}  & [0.158, 0.195]    & {\bf -1.937} \\
       & Hol        &-0.013  & -1.472  & [0.149, 0.177]    & [-1.268, -1.230]  & [0.204, 0.242]    & -1.246 \\
\hline
\label{tab:dipolecorrection}
\end{tabular}
\end{table}

\clearpage

\clearpage

\clearpage

\clearpage

\clearpage

\clearpage

\clearpage


\end{document}